# Visualization of spin-polarized electronic states by imaging-type spin-resolved photoemission microscopy

Koichiro Yaji* and Shunsuke Tsuda

*Center for Basic Research on Materials, National Institute for Materials Science, 3-13 Sakura, Tsukuba, Ibaraki, 305-0003, Japan*

*Corresponding author: yaji.koichiro@nims.go.jp

Abstract

Harnessing electron spin is crucial in developing energy-saving and high-speed devices for the next generation. In this scheme, visualizing spin-polarized electronic states aids in designing and developing new materials and devices. Spin-resolved photoemission spectroscopy provides information on the spin-polarized electronic states. To investigate the spin-polarized electronic states in microscopic materials and devices, spin-resolved photoemission spectroscopy requires spatial resolution in a sub-micrometer scale. Here we show the imaging-type spin-resolved photoemission microscopy (iSPEM) with an ultraviolet laser developed at the National Institutes for Materials Science (NIMS). Our iSPEM achieves a spatial resolution of 420 nm, drastically improving by more than an order of magnitude compared to conventional spin-resolved photoemission spectroscopy instruments. Besides, the multi-channel spin detector significantly reduces the data acquisition time by four orders of magnitude compared to the conventional instruments. The iSPEM machine elucidates the spin-polarized electronic states of sub-micrometer scale materials, polycrystals, device structure samples, and so on, which have yet to be the target of conventional spin-resolved photoemission spectroscopy.

*1. Introduction*

The convergence of spintronics, topological physics, and quantum technology rapidly expands the potential for innovative changes in modern science and technology. These fields have developed independently, yet merging allows for new approaches. For example, topological spintronics aims to improve information processing and device technology by utilizing electron spins and topological properties [1–3]. Quantum spintronics enables ultrafast quantum computing by controlling the spins in a qubit [4,5]. Using electron spin permits devices with high speed and low power consumption in these technologies.

The spin-polarized electronic bands characterize the properties of functional materials utilizing the spins. Therefore, visualization of spin-polarized electronic states is helpful for the development of new materials and devices. In addition, quantitative evaluation of both spin polarization and spin orientation is essential for improving the performance of the materials and the devices. Spin and angle-resolved photoemission spectroscopy (SARPES) is a powerful technique that can independently observe the energy, momentum, and spin of electrons [6]. Here, conventional electronic structure studies using SARPES have been dominated by measurements of spin-polarized band structures of ideal single crystals with sizes typically ranging from several hundred micrometers to several millimeters. SARPES for such ideal systems has significantly contributed to the progress in condensed-matter physics and electronic properties [7,8,9].

The development of SARPES instruments has been made a great effort, driven by the growing demand to investigate the spin-polarized electronic states of novel materials [10]. In SARPES, the energy and momentum of the photoelectron are examined using a photoelectron analyzer first. Subsequently, the photoelectron is led to the spin detector for spin analysis. Before 2000, SARPES utilized a Mott-type spin detector [11]. The efficiency of the spin detection with the Mott detector was extremely low, necessitating considerable time to acquire the data. The development of a very-low-energy-electron-diffraction (VLEED) type spin detector in the 21st century significantly improved the efficiency of SARPES measurements [12], and SARPES has been widely employed in the study of electronic properties. Besides, the combination of a large hemispherical photoelectron analyzer, the VLEED spin detector, and a high-brilliant vacuum ultraviolet laser realized the high resolution of the SARPES (laser-SARPES), where the energy resolution of 1 meV and a wavenumber resolution of 0.01 Å$^{-1}$ were achieved [13]. The laser-SARPES has significantly advanced the understanding of the spin-polarized electronic states of topological materials [14-20] and the spin interference in the photoexcitation process [21,22]. In addition to improvements in spin filters, tremendous efforts have been made to improve data acquisition methods in the spin detector. The conventional spin detectors employ a single-channel method, in which photoelectrons of an energy ($E$) and a momentum ($k_x$, $k_y$) are led to the spin detector, and their spin information is analyzed. Thus, spin-polarized band mapping over the whole Brillouin zone requires enormous time. This situation makes it difficult to systematically investigate the spin-polarized electronic states in various materials. Recently, a multichannel spin detector based on an imaging concept has been developed, allowing for much more efficient data acquisition in SARPES. The multichannel spin detector analyzes the spins of an $E$–$k_x$ or $k_x$–$k_y$ image obtained by the two-dimensional photoelectron analyzer simultaneously, of which the measurement efficiency can be a factor of four compared with the single-channel method [23]. As the multichannel spin detectors, spin-polarized low-energy electron scattering (SPLEED) type using W(001) or Au/Ir(001) surfaces [23–26], the VLEED type using an O/Fe(001) surface [27], and an imaging Mott type have been reported

[28]. The VLEED-type spin detectors use exchange scattering on ferromagnetic surfaces to distinguish the spins. In contrast, the SPLEED-type spin detectors use spin-dependent diffraction intensity differences on heavy-element surfaces to identify the spins.

Photoemission microscopy offers tremendous opportunities for characterizing the electronic structure of microscopic materials and devices. Spin-integrated photoemission microscopy has been successfully performed at several synchrotron radiation facilities, allowing the measurement of valence bands with sub-micrometer spatial resolution [29–32]. We have recently developed a spin-integrated photoemission microscopy apparatus incorporating a vacuum-ultraviolet laser as a lab-based machine [33]. In the spin-resolved measurements, micro-SARPES has been carried out using a laser focused to 10–50 micrometers [13, 34]. Using the micro-SARPES, a weak topological insulator phase in bismuth iodate [15] and a higher-order topological insulator phase of bismuth bromide chains [17] have been successfully demonstrated. In general, the sizes of quantum and spintronics materials and devices are less than a few micrometers. Therefore, the spatial resolution of the spin-resolved photoemission microscopy requires less than the sub-micrometers to visualize the spin-polarized electronic state of the microscopic materials and devices. However, such a machine has not been realized so far.

In this article, we report on imaging-type spin-resolved photoemission microscopy (iSPEM) with high spatial resolution developed at the National Institute for Materials Science (NIMS). Our iSPEM machine is equipped with state-of-the-art photoemission microscopy technology and highly efficient spin-resolved data acquisition capability using the multi-channel spin detector. We show the spatially resolved spin polarization imaging of the valence states of a polycrystalline iron (poly-

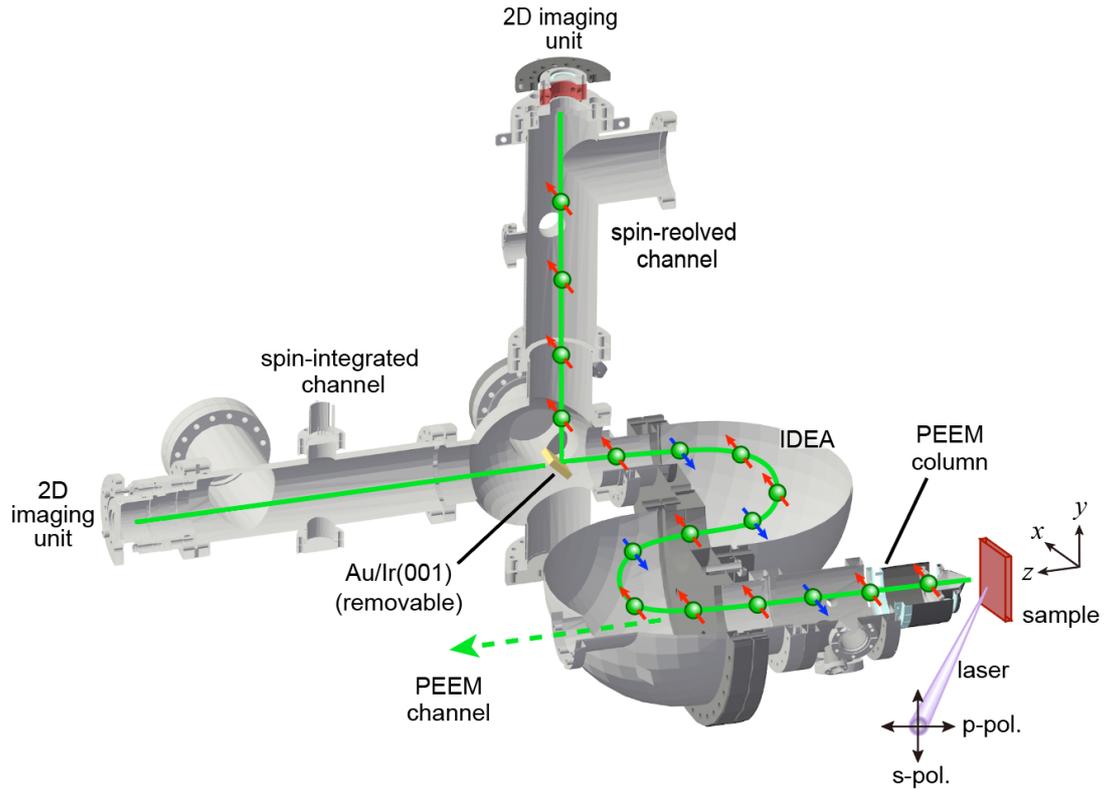

Figure 1. Overview of the iSPEM instrument with the multichannel spin detector. Green lines represent the photoelectron trajectories.

Fe) in real-space measurement mode. The spatial resolution of the spin-resolved photoemission microscopy is 420 nm. As a demonstration in momentum space measurement mode, we show the quick imaging of the spin-resolved Fermi surface in the whole Brillouin zone of the bismuth (Bi) (111) surface.

## 2. Instruments

The iSPEM machine is based on the apparatus reported in reference 33, which consists of three ultra-high vacuum chambers: an analysis chamber, an intermediate chamber, and a sample preparation chamber. The analysis chamber is equipped with a vacuum ultraviolet laser ($h\nu = 10.9$ eV), a helium discharge lamp (He lamp) ($h\nu = 21.2$, 40.8 eV), and a mercury vapor lamp (Hg lamp) ($h\nu = 5.2$ eV) as excitation light sources.

Nano-ESCA made by FOCUS GmbH is employed for photoelectron analysis (Fig. 1). Photoelectrons emitted from the sample are accelerated to 12–20 kV by the electrostatic field between the sample and the first lens of the analyzer. Subsequently, the photoelectrons enter the photoelectron emission microscopy (PEEM) column. This PEEM lens system realizes photoelectron microscopy with high spatial resolution. In addition, real-space ($x$–$y$) imaging and momentum-space ($k_x$–$k_y$) imaging can be easily switched thanks to the PEEM lens system. Photoelectrons passing through the PEEM column are energy-filtered by an Imaging Double Energy Analyzer (IDEA). An iridium (001) single crystal passivated by a monolayer gold [Au/Ir(001)] is installed at the exit lens section of IDEA. Photoelectrons passing through IDEA are guided to the Au/Ir(001) crystal for spin analysis, where the two-dimensional $x$–$y$ or $k_x$–$k_y$ information is preserved before and after scattering by the Au/Ir(001) crystal. In our setup, the spin detector is arranged to detect the $x$-component of the spin polarization. A spin-integrated image can be obtained by retracting the Au/Ir(001) crystal. The photoelectrons scattered by the Au/Ir(001) crystal are detected by a two-dimensional imaging unit. Here, we note that Nano-ESCA produces the $x$–$y$ image in real-space mode and the $k_x$–$k_y$ image in momentum-space mode, while the ARPES measurement using a typical hemispherical photoelectron analyzer acquires an $E$–$k_x$ image. Nano-ESCA can simultaneously measure electronic states within $k_{x,y} \sim \pm 6.2$ Å$^{-1}$, significantly reducing the data acquisition time compared with conventional ARPES instruments. Nano-ESCA eliminated the sample rotation for angle-resolved measurements in contrast to the conventional ARPES and SARPES. This capability provides a significant advantage in the measurements of tiny samples.

The iSPEM machine adopts the SPLEED-type spin detector using the Au/Ir(001) crystal. In this spin detector, the photoelectrons are reflected by the

Au/Ir(001) crystal with the scattering energy at the working points (WP$_1$ and WP$_2$). The photoelectron spins are distinguished using the asymmetry of the spin-dependent scattering intensity at WP$_1$ and WP$_2$. This corresponds to observing the (0 0) spot intensity in low energy electron diffraction (LEED). The exit lens section of IDEA adjusts the energy of the photoelectron incident on the Au/Ir(001) crystal while keeping the image information. The details of the spin detector using the Au/Ir(001) crystal are described in the literature [35, 36].

We describe how to extract the unknown spin polarization $P$ of the photoelectron from the experimentally measured scattering intensity $I_1$ and $I_2$ at WP$_1$ and WP$_2$ [25, 37]. The following equations give the intensities $I_1$ and $I_2$:

$$I_1 = I_0 R_1 (1 + S_1 P) \qquad (1)$$

$$I_2 = I_0 R_2 (1 + S_2 P), \qquad (2)$$

where $I_0$ is the unknown photoelectron intensity incident to the Au/Ir(001) crystal, the $R_1$ and $R_2$ are reflectivity at WP$_1$ and WP$_2$, respectively. The values of $R_1$ and $R_2$ are experimentally evaluated using spin unpolarized photoelectrons emitted from a silver polycrystal and are determined to be 0.7 % and 2.1 %, respectively. $S_1$ and $S_2$ are the spin sensitivities at WP$_1$ and WP$_2$, the so-called effective Sherman functions. The manufacturing test evaluates these values as $S_1 = -0.68$ and $S_2 = 0.55$. From the simultaneous equations (1) and (2), $P$ and $I_0$ are obtained as

$$P = \frac{I_1/R_1 - I_2/R_2}{S_1 I_2/R_2 - S_2 I_1/R_1} \qquad (3)$$

$$I_0 = \frac{S_2 I_1/R_1 - S_1 I_2/R_2}{S_2 - S_1}. \qquad (4)$$

By using (3) and (4), the spin-resolved photoelectron images $I_\uparrow$ and $I_\downarrow$ are given by

$$I_\uparrow = I_0 (1 + P) \qquad (5)$$

$$I_\downarrow = I_0 (1 - P). \qquad (6)$$

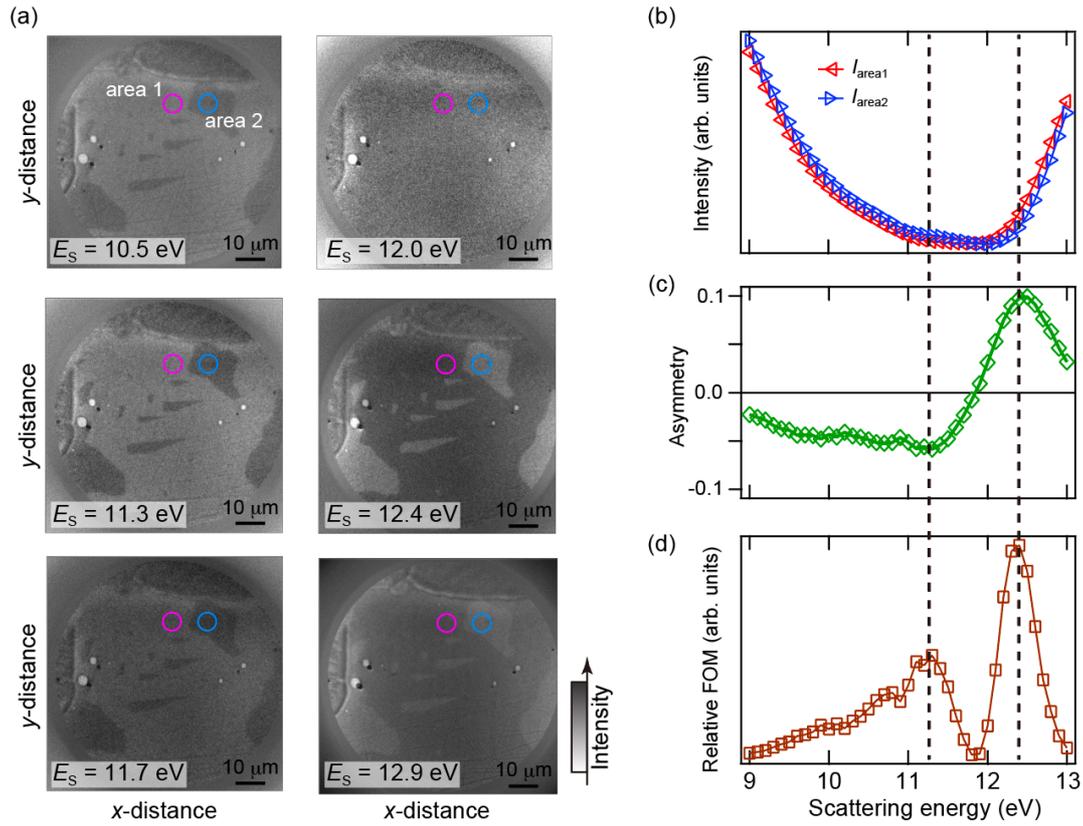

Figure 2. (a) Photoemission microscopy images of poly-Fe scattered by the Au/Ir(001) crystal, where only images at selected scattering energies are shown. (b) IV curves, where $I_{area1}$ and $I_{area2}$ are obtained by the summation of the intensity within the area 1 and area 2 shown in (a), respectively. The asymmetry (c) and the relative FOM (d) obtained by the formulas (7) and (8) are displayed.

We note that $I_1$ and $I_2$ obtained by the experiments does not directly provide spin-resolved electronic states.

## 3. Demonstration

### 3-1. Spin-resolved photoemission microscopy in real space

We show spin-polarized valence state imaging of poly-Fe in the real space mode to characterize our iSPEM. The clean surface of poly-Fe was obtained by repeated cycles of $Ar^+$ sputtering and annealing at 500°C. The Hg lamp was used as an excitation light source. The iSPEM measurements were performed with a field of view (FoV) of 80 μm with the binding energy $E_B$ = 0.4 eV at room temperature.

In Fig. 2(a), we show the scattering energy ($E_s$) dependence of the photoemission microscopy images by the Au/Ir(001) crystal, where only typical examples are displayed. The measurements were performed from the 9.0 eV to 13.0 eV scattering energies with 0.1 eV energy steps. Each image was recorded with a dwell time of 10 sec. In the images of $E_s$ = 10.5, 11.3, 11.7 eV, there are several dark gray areas in the light gray area. Light and dark gray areas are not discernible in the image obtained with $E_s$ = 12.0 eV. In contrast, the light and dark gray areas in the images of $E_s$ = 12.4 and 12.9 eV are reversed. This color contrast reflects the $x$-component of the spin polarization of magnetic domains in poly-Fe.

We determine the scattering energies ($E_{s1}$ and $E_{s2}$) for spin detection. Figure 2(b) shows the photoelectron intensities at area 1 and area 2 ($I_{area1}$ and $I_{area2}$) as a function of $E_s$ (IV curves). From the IV curves, the asymmetry ($Asym$) is calculated by

$$Asym = \frac{I_{area1} - I_{area2}}{I_{area1} + I_{area2}} \qquad (7)$$

[Fig. 2(c)]. Here, we define the relative figure of merit (FOM) as follows:

$$FOM = (Asym)^2 \cdot (I_{area1} + I_{area2}). \qquad (8)$$

From the FOM shown in Fig. 2(d), the working points (WP$_1$ and WP$_2$) for the spin detection are estimated to be $E_{s1}$ = 11.2 eV and $E_{s2}$ = 12.4 eV, respectively.

Figure 3(a) shows a spin-resolved photoemission microscopy image of poly-Fe using the data taken at WP$_1$ and WP$_2$. Here, the absolute value of the spin polarization cannot be determined since the magnetization direction of each magnetic domain of poly-Fe is unknown. By defining the spin quantization axis of the sample, the spin polarization projected on the $x$-axis can be evaluated quantitatively. The spatial resolution of our iSPEM is evaluated from the profiles along #1 and #2 in Fig. 3(a) [Fig.

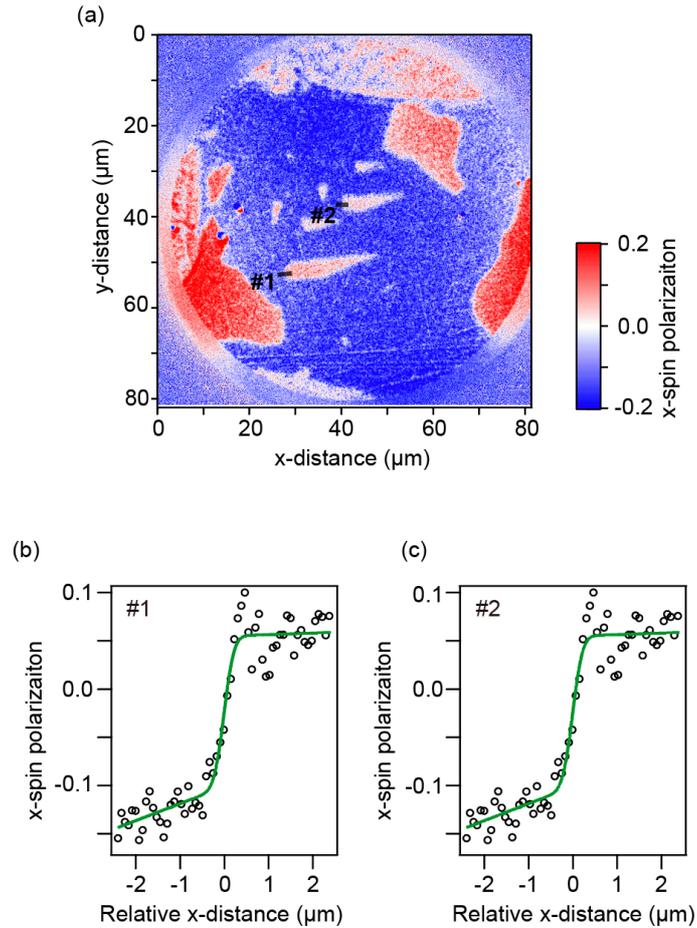

Figure 3. (a) Spin polarization imaging of poly-Fe. The *x*-spin component is detected. The spin polarization reflecting the magnetic domains in poly-Fe is visualized. Line profiles along #1 (b) and #2 (c) are displayed. Circles give the experimental data. Solid curves represent the fitting results by the step function convoluted with the Gaussian.

3(b,c)]. The profiles are fitted by a step function convoluted with a Gaussian. We defined the spatial resolution by the full width at half maximum of the Gaussian. As a result, the spatial resolution at #1 (#2) is estimated to be 420 nm. The spatial resolution of this system in spin-integrated photoemission microscopy is 30 nm [33]. The spatial resolution in spin-resolved mode is an order of magnitude lower than that in spin-integrated mode. One reason for this is the complexity of adjusting the electronic lens system around the Au/Ir(001) crystal at $WP_1$ and $WP_2$, resulting in defocusing the image after reflection by the Au/Ir(001) crystal. Another possible reason is that the vibration of the instrument affects the spatial resolution of the spin-resolved measurement because the data acquisition time for the spin-resolved measurement is much longer than that for the spin-integrated measurements.

### *3-2. Spin-resolved photoemission spectroscopy in momentum space*

We show a Fermi surface mapping of a Bi(111) single-crystal film by ARPES and SARPES with the vacuum ultraviolet laser. The Bi(111) surface states are spin-polarized due to strong spin-orbit interaction [38, 39], making it suitable for the demonstration of SARPES. The Bi(111) single-crystal film was *in situ* grown on the Ge(111) substrate. The clean surface of the Ge(111) substrate was obtained by repeated cycles of $Ar^+$ sputtering and annealing up to 600 ºC. Subsequently, a 100-bilayer Bi film was deposited on the Ge(111) substrate at room temperature. The ARPES and SARPES measurements were performed with the experimental geometry shown in Fig. 4(a). The photoelectrons were excited by the 10.9-eV photons. The light-incident plane is defined as the plane spanned by the light-incident axis and the normal axis of the sample surface. In the present study, we arranged the $\bar{\Gamma}\bar{M}$ mirror plane of the sample perpendicular to the light incident plane. The electric field vector of the laser was parallel to the light incident plane (p-polarized light). The sample temperature was kept to 40 K during the measurements. In SARPES, we observed the *x*-component of the spin polarization. The real space FoV was set to 46 μm.

Figure 4(b) shows the ARPES intensity mapping at the Fermi level of Bi(111). The data acquisition time is 10 sec. Our instrument enables the Fermi surface mapping of the whole Brillouin zone in a single measurement. In Fig. 4(b), a closed Fermi surface ($S_1$) centered at the $\bar{\Gamma}$ point, elliptical Fermi surfaces ($S_2$) and a needle-like Fermi surface ($S_3$) elongated in the $\bar{\Gamma}\bar{M}$ direction are observed. These $S_1$-$S_3$ are the surface states of Bi(111), being in agreement with the previous studies [40, 41].

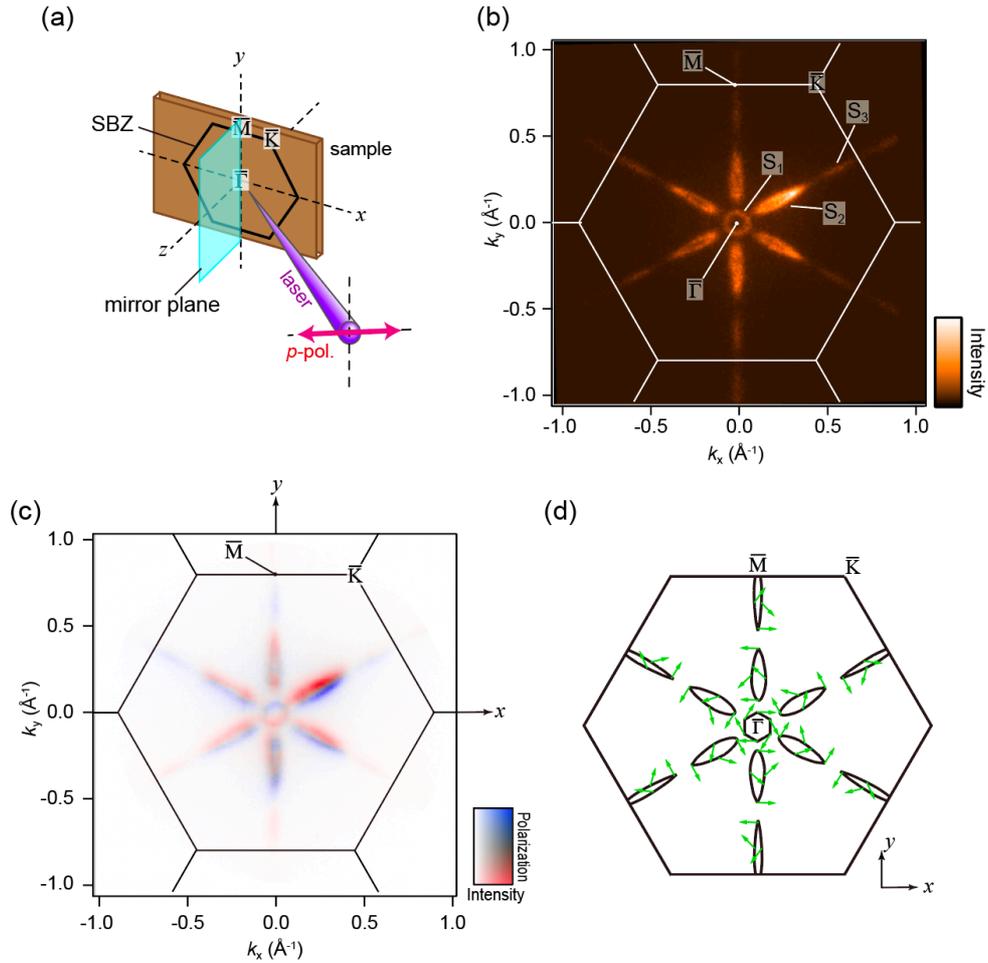

Figure 4. (a) Schematic drawing of the experimental geometry. Light blue parallelogram represents the mirror plane of crystal, which is perpendicular to both the light incident plane and the electric field vector of the light. Spin-integrated (b) and spin-resolved (c) Fermi surface mappings of Bi(111) are shown. Thin solid lines give edges of the surface Brillouin zone. (d) Spin texture of the spin-polarized surface states for Bi(111) based on the theoretical calculations [45,46]. The arrows represent the spin direction projected in the plane. Here, the information of the spin polarization value is omitted in this figure.

In an ideal model of the spin texture of the spin-orbit coupled surface states, the spin is oriented perpendicular to both the electron momentum and the surface normal [42,43], the so-called spin-momentum locking. Our iSPEM is arranged for the detection of the $x$-spin polarization. Thus, to observe the spin polarization of the Bi(111) Fermi surfaces elongated to the $\overline{\Gamma M}$ direction, the SARPES measurements were performed with the experimental geometry that the $\overline{\Gamma M}$ axis is parallel to the *y*-axis. Here, the Bi(111) surface has three-fold rotational symmetry, so that one of the three equivalent $\overline{\Gamma M}$s is parallel to the *y*-axis as shown in Fig. 4(a). Figure 4(c) exhibits the spin-resolved Fermi surface mapping. In the $\overline{\Gamma M}$ direction on the $k_y$-axis, the spin polarizations of $S_1$, $S_2$, and $S_3$ are inverted with respect to $k_y = 0$ due to the time-reversal symmetry. On the other hand, in the $\overline{\Gamma M}$ direction rotated by ±60º from the *y*-axis, the *x*-component of the spin polarization of $S_2$ is reversed with respect to the $\overline{\Gamma M}$ line. The ideal model of the spin-momentum locking cannot explain the observed spin polarization for $S_2$. According to recent experimental and theoretical studies, the spin texture of spin-orbit coupled surface states depends on the symmetry of the surface structure and the details of the electronic orbitals [44]. We refer to theoretical calculations of the spin texture for the Bi(111) surface states to understand our experimental results [Fig. 4(d)] [45,46]. The spin of $S_2$ in the $\overline{\Gamma M}$ mirror plane is exactly pointing in the direction perpendicular to the mirror plane. The general description of the spin-orbital texture in the mirror plane is described in the reference [47]. On the other hand, the spins are pointing in a different direction from that predicted by the spin-momentum locking model away from the $\overline{\Gamma M}$ mirror plane. The experimental results shown in Fig. 4(c) are explained by the projection of the calculated spin texture onto the *x*-axis.

The *x*-component of the spin polarization observed in the SARPES measurements qualitatively agrees with theoretical calculations. Here, one should be careful in quantitatively treating the spin polarization of the spin-orbit coupled surface state observed by SARPES, i.e., the direction of the spin vector and the value of the spin polarization. The photoemission process is described by a dipole transition matrix element $\langle \psi_f | \mathbf{A} \cdot \mathbf{p} | \psi_i \rangle$, where $|\psi_i\rangle$ is the initial state, $|\psi_f\rangle$ the final state, $\mathbf{A}$ the vector

potential of the light, and **p** the electron momentum operator. The matrix element contains information on not only the initial state but also on the photoexcitation process. Therefore, the spin polarization of photoelectrons observed by SARPES can be modified due to the matrix element effect, which is systematically studied in the previous studies [47-50].

In the above demonstration, we set the real-space FoV of ARPES and SARPES to 46 μm. The magnification of the real-space FoV can be optimized from a few micrometers to several hundred micrometers, depending on the area of interest. In addition, the SARPES measurements in sub-micrometer scale (nano-SARPES) are available using a spot selector (continuously variable iris aperture).

## *4. Summary and outlook*

We have developed imaging-type spin-resolved photoemission microscopy (iSPEM) with a 10.9-eV laser. Thanks to the multichannel spin detector, the data acquisition efficiency in spin-resolved photoemission spectroscopy has been improved by four orders of magnitude compared with the conventional machines using the single-channel one. The PEEM lens system has realized the high spatial resolution in iSPEM. In the real-space mode of iSPEM, the spatial resolution of 420 nm has been achieved, more than an order of magnitude higher than that of the conventional spin-resolved photoemission spectroscopy. In the momentum-space mode of iSPEM, the spin-resolved Fermi surface imaging of Bi(111) has been demonstrated.

The highly efficient spin detection capability and the high spatial resolution of iSPEM enable the observation of the spin-polarized electronic states of tiny samples and inhomogeneous samples, for instance, polycrystalline, powder, and combinatorial samples, which are challenging to measure the electronic states by the conventional spin-resolved photoemission spectroscopy. Besides, visualization of the spin polarization of functional electrons is significant for designing and developing spin-based devices. The iSPEM technique would dramatically advance the study of electronic properties in various materials, such as quantum, magnetic, and low-dimensional materials, as well as in devices utilizing these materials.

Acknowledgements

The authors thank Nils Weber for technical support in developing the spectrometer. The authors also thank Shingo Takezawa for technical support in preparing the Bi/Ge(111) sample.

Funding

The present work was partially supported by the Japan Society for the Promotion of Science KAKENHI (Grant Nos. JP21K04633 and JP22H01761), the Innovative Science and Technology Initiative for Security Grant Number JPJ004596, ATLA, Japan, and Iketani Science and Technology Foundation.

Disclosure statement
The authors report there are no competing interests to declare.